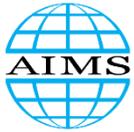
*Public Health*

*Research article*

# Automatic COVID-19 disease diagnosis using 1D convolutional neural network and augmentation with human respiratory sound based on parameters: cough, breath, and voice

**Kranthi Kumar Lella and Alphonse Pja\***

Department of Computer Applications, NIT Tiruchirappalli, Tamil Nadu, India

**\* Correspondence:** Email: alphonse@nitt.edu; Tel: +91(431)2503742; Fax: +91(431)2500133.

**Abstract:** The issue in respiratory sound classification has attained good attention from the clinical scientists and medical researcher's group in the last year to diagnosing COVID-19 disease. To date, various models of Artificial Intelligence (AI) entered into the real-world to detect the COVID-19 disease from human-generated sounds such as voice/speech, cough, and breath. The Convolutional Neural Network (CNN) model is implemented for solving a lot of real-world problems on machines based on Artificial Intelligence (AI). In this context, one dimension (1D) CNN is suggested and implemented to diagnose respiratory diseases of COVID-19 from human respiratory sounds such as a voice, cough, and breath. An augmentation-based mechanism is applied to improve the preprocessing performance of the COVID-19 sounds dataset and to automate COVID-19 disease diagnosis using the 1D convolutional network. Furthermore, a DDAE (Data De-noising Auto Encoder) technique is used to generate deep sound features such as the input function to the 1D CNN instead of adopting the standard input of MFCC (Mel-frequency cepstral coefficient), and it is performed better accuracy and performance than previous models. *Results:* As a result, around 4% accuracy is achieved than traditional MFCC. We have classified COVID-19 sounds, asthma sounds, and regular healthy sounds using a 1D CNN classifier and shown around 90% accuracy to detect the COVID-19 disease from respiratory sounds. *Conclusion:* A Data De-noising Auto Encoder (DDAE) was adopted to extract the acoustic sound signals in-depth features instead of traditional MFCC. The proposed model improves efficiently to classify COVID-19 sounds for detecting COVID-19 positive symptoms.

**Abbreviations:** CNN: Convolutional Neural Network; DDAE: Data De-Noising Auto Encoder; MFCC: Mel-frequency Cepstral Coefficient; DL: Deep Learning; ML: Machine Learning; AI: Artificial Intelligence; SVM: Support Vector Machine; LVQ: Learning Vector Quantization; MLR: Multivariate Linear Regression; MRI: Magnetic Resonance Imaging; SSP: Speech Signal Processing; LSTM: Long Short-Term Memory; TDSN: Tensor Deep Stacking Network; CRD: Compression of Range Dynamically; BN: Background Noise; ST: Stretching Time; SP: Shift Pitch; ReLU: Rectified Linear Unit; MUDA: Musical Data Augmentation; JAMS: JSON Annotated Music Specification;

## 1. Introduction

As of 23rd January 2021, the COVID-19 epidemic is declared as a pandemic by the World Health Organization (WHO) on March 11th, 2020, and it claims over 2,098,879 lives worldwide [1]. As of 23rd January 2021, a global situation was confirmed with 2,098,879 cases of COVID-19, including 2,098,879 deaths as shown in Figure 1. Experts in microbiology believe that data collection is critical for isolating infected people, tracing connections, and slowing the spread of the virus. Although progress in testing has made these methods more popular in recent months, it is imperative to have affordable, simple, and scalable COVID-19 screening technologies. The seriousness of COVID-19 disease is classified into three categories: extreme, middle/moderate, and mild. The problem of respiratory sound classification [2,3] and diagnosis of COVID-19 disease has received good attention from the clinical scientists and researchers community in the last year. In this situation, many AI-based models [4–6] entered into the real-world to solve such problems; and researchers have provided different machine learning, signal processing, and deep learning techniques to solve the real-world problem [7,8].

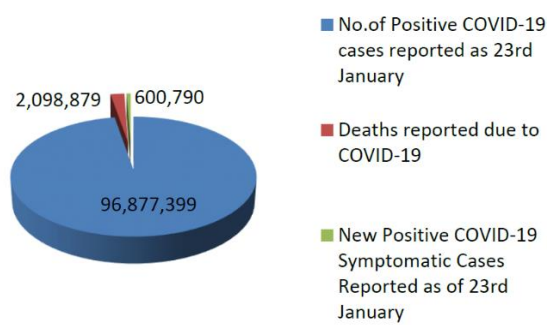

**Figure 1.** The COVID-19 situation reported to WHO (World Health Organization) as of 23rd January 2021.

Nowadays, the COVID-19 pandemic exists in the entire real-world and generates fear in communicating physically. There are several ways to diagnose COVID-19 disease, and one of them is through human respiratory sounds. Clinical experts use respiratory sounds generated by the human body like vibration, voice, lung sound, heart, food absorption, breathing, cough, and sighs to diagnose the disease [9]. To date, such signals are typically obtained during scheduled visits via manual auscultation. Technological researchers and medical scientists have now begun using electronic technologies to collect sound from the human body (like Digital- Stethoscopes) [10] and carry out an automated examination of the human sounds data, e.g., recognition of wheeze in asthma. Researchers pilot the use of the human voice to aid early detection of several diseases: Alzheimer's disease corresponds with normal to slur [11], stammer, repeat, and use incomplete phrases and words (The beloved one can have trouble forming clear phrases or recognizing conversations), Parkinson's Disease (PD) may have many effects on the voice (several people with PD talk softly and they don't show enough feeling in one tone, speaking voice breathy or hoarse occasionally, and the end of a phrase, people with Parkinson's could slur sentences, mumble or trail off) [12,13], frequency of speech with coronary heart disease (people can develop neck pain, fatigue, voice disorder) [14,15], invisible conditions such as battle fatigue, brain trauma, and psychological situations correlate with sound pitch, vocal tone, speech rhythm & frequency, and voice sound volume. The diagnostic tool for various diseases is the use of human respiratory sound. It provides enormous potential for early detection and cheap solutions for products rolling into the masses. It is valid for people that the solution can be tracked in their everyday lives by various people. The efficiency of the respiratory sound classification COVID-19 sounds dataset is enhanced in the last year by applying separate ML (Machine Learning) techniques [16] such as SVM (Support Vector Machine), LVQ (Learning Vector Quantization), and MLR (Multivariate Linear Regression).

A one-dimensional CNN model is currently proposed and applied to classify the human sound signs based on respiratory signs in familiar and unusual ways, regardless of cough, voice, and breathing. In addition, a Data De-noising Auto Encoder (DDAE) algorithm is used as an input instead of following the traditional MFCC (Mel-frequency Cepstral Coefficient) to obtain depth features of respiratory sounds/voice, which is the basic input function to the 1D convolution. The functionality of the 1D Convolutional Neural Network (1D CNN) subsequently transformed the depth input features derived by the DDAE (Data De-noising Auto Encoder) method and then executed pooling activity. Finally, using a 'Softmax' classifier, the processed signals are categorized. The DDAE is derived in-depth feature respiratory sound signals in contrast with the standard MFCC feature extraction method [17–19]; the usefulness of the fundamental function of this analysis throughout the classification of respiratory sounds is thus demonstrated. The accuracy of classification with 1D CNN has a high '$F_1$' Score which for the diagnosis of COVID-19 is slightly increased. An electronic/digital stethoscope can obtain the human respiratory sounds; the presented model is applicable and demonstrates good robustness [13,16].

Recent research has begun to investigate how the respiratory sounds recorded by smart devices from patients who have confirmed positive cases of COVID-19 in the clinic vary with the respiratory sound signs of healthy people (e.g., breath sound, cough sound, and voice) [20,21]. Lung auscultation digital stethoscope data can be used for diagnosis of COVID-19 disease; a COVID-19 related cough detection analysis obtained with iOS/Android phones is presented using a group of forty-eight (48) patients with COVID-19 symptoms other clinical coughs trained in a series of models. Until then, it is only possible for the system to automatically classify the patient's health fitness and analyze the symptoms of the affected patient's respiratory sounds from clinical patients. It is not the case in our work, which includes studying human respiratory sounds in crowdsourced, unregulated data for the diagnosis of COVID-19 disease.

Various COVID-19 sounds in the name of "COVID-19 crowdsourced sounds data" are collected by us from Cambridge University with mutual agreement. The dataset is obtained through an android/iOS mobile app and internet in the form of voluntary speech, cough, and respiratory and clinical samples with background and signs. This iOS/Android app also collects information on whether the user is tested with a positive COVID-19 case before or not. To date, this COVID-19 sounds dataset has collected around 8000 unique users in order of 12,000. Although there are other attempts to gather user-related data, they are mostly either narrow in reach or size. To our understanding, this is the world's most significant unregulated, crowdsourced data set of COVID-19 based sounds. All these data are collected from Cambridge University with mutual agreement. This is a specific reason to collect the COVID-19 sounds dataset and perform 1D convolutional with augmentation method to diagnose the disease as well. In this paper, we briefly described the COVID-19 data, COVID-19 Sound Analysis, and proposed 1D CNN approach in Section 1; Section 2 illustrates literature reviews of each author and background study for this research; Section 3 describes the proposed 1D convolutional method, dataset collection, and augmentation process; Section 4 outlines the result analysis and discussions concerning the proposed model, and we have concluded this research with the best accuracy.

## 2. Background works

Researchers and scientists have long recognized the utility of sound as a potential predictor of actions and health. For example, independent audio recorders [22] were used for the reason in digital stethoscopes to identify sounds from the human respiratory. Highly trained clinicians felt that compared to other methodologies such as MRI (Magnetic Resonance Imaging), and sonography, it was effortless to listen to sounds and then interpret them. However, recent work on automated sound modeling and interpretation has the potential to deal with these methods and provide the respiratory sound as an alternative that is relatively inexpensive and easily distributed. The microphones were recently exploited for sound processing on goods and product-based machines such as android/iOS devices (smartphones) and wearables technologies.

In [23], Brown C et al. proposed an Android/iOS app to collect COVID-19 sounds data from crowdsourced sounds respiratory data of more than 200 positives for COVID-19 from more than 7k unique users; Brown C et al. used general parameters and 3 major set COVID-19 tasks based on breath and cough sound. The parameters are: i) positive COVID-19/negative COVID-19; ii) positive COVID-19 with cough/negative COVID-19 with cough; iii) positive COVID-19 with cough/non-COVID asthma cough. Task 1 achieved 80% accuracy with 220 on dealing with the combination of cough and breath; task 2 achieved 82% of accuracy with 29 users on dealing with a cough only; finally, task 3 achieved 80% accuracy with 18 users on dealing breath. Recall function is slightly low (72%) because of not having a specialized net to detect every COVID-19 cough. Brown C et al. used an SVM classifier to analyze the sound signals. In this case, instead of SVM, we proposed a 1D CNN and augmentation approach with a data de-noising method to classify and diagnose the COVID-19 disease.

In [24], Han J et al. conducted an intelligent analysis of COVID-19 speech data by considering four parameters, such as: i. Sleep Quality; ii. Severity; iii. Anxiety; iv. Fatigue. Kun Q et al. collected data from the "COVID-19 sounds app" launched by scientists and researchers from Cambridge University and the "Corona voice detect App" launched by researchers from Mellon University. After data processing, people have obtained 378 total segments. From this preliminary study, they have taken 260 recordings for future analysis. These 256 sound pieces were collected from 50 COVID-19 infected patients. For future study, poly impulses with a sample rate of 0.016MHz were converted. They have considered two acoustic feature sets in this study, namely ComParE & eGeMAPS. Both feature sets achieved 69% accuracy. In this paper, the authors have taken COVID-19 affected patient speech data to classify sleep quality, severity, anxiety, and fatigue. However, in our research, we used different respiratory sounds to train the disease model and diagnosis.

In [25], Orlandic L et al. implemented the "COUGHVID" crowdsourced dataset for cough analysis with COVID-19 symptom; More than twenty thousand crowdsourced cough recordings reflected a broad range of topic gender, age, geographic locations, and COVID-19 status was given in the COUGHVID dataset. They have collected a series of 121 cough sounds and 94 no-cough sounds first-hand to train the classifier includes voice, laughter, silence, and various background noises [26]. They have taken self-reported status variables (25% of recording sounds with healthy values, 25% sound recordings with COVID values, 35% sound recordings with symptomatic value, and 15% sounds recordings with non-reported status; It ensured that all three reviewers labeled 15% of cough sounds for the selection of the recordings. The percentage of COVID positive symptoms of COVID-19 and healthy subjects were 7.5%, 15.5, and 77% from 65.5% males and 34.5% females, respectively. In this paper, the authors have collected cough sounds with the name "COUGHVID" and detected COVID-19 from cough sounds only. In the present approach, cough, breath, and continuous speech are considered for analysis COVID-19 disease identification.

In [2], Wang Y et al. proposed a method to classify large-scale screening of people getting infected with COVID-19. This work was to identify various breathing patterns. In this paper, first, a new and robust RS (Respiratory Simulation) Model is introduced to fill the gap between a massive amount of training data and inadequate actual data from the real-world to consider the features of accurate respiratory signals. To identify six clinically significant respiratory patterns, they initially applied bidirectional neural networks like the GRU network attentional tool (BI_at_GRU) (Tachypnea, Eupnea, Biots, Cheyne-Stokes, Bradypnea, and Central-Apnea). In comparative studies, the acquired BI_at_GRU specific to the classification of respiratory patterns outperforms the existing state-of-the-art models. The proposed deep model and design concepts have enormous potential to be applied to large-scale applications such as sleeping situations, public environments, and the working environment. In this paper, the authors have collected different breathing patterns to detect the disease from collected respiratory sounds. However, in the proposed approach, three (breathing patterns, voice sound patterns, cough sound patterns) and a combination of all respiratory sounds are collected and analyzed.

In [3], a portable non-contact system was proposed by Jiang Z et al. to track the health status of individuals wearing masks by examining features of the respiratory system. This device consists of mainly a thermal imaging camera with FLIR (Forward-looking Infrared) and an Android device. Under realistic situations such as pre-screening in institutions and clinical centers, this can help distinguish those possible COVID-19 patients. In this work, they performed health screening using thermal and RGB videos from DL architecture-based cameras. Firstly, they used pulmonary data analysis techniques to recognize mask-wearing people to obtain the health screening outcome, and a BI_at_GRU function is applied to pulmonary disease results. As a result, 83.7% accuracy is achieved to classify the respiratory health conditions of a diseased patient. In this, thermal images and breathing patterns were collected to predict the COVID-19 disease symptoms, but in the current system, only three significant respiratory sounds related to COVID are collected for detecting the COVID-19 disease.

In [4], Imran A et al. implemented an AI (Artificial Intelligence) based screening solution to detect COVID, transferable through a smart mobile phone application. The application was suggested, developed, and finally tested. The mobile app called AI4COVID-19 records and sends to an AI-based cloud running in the cloud triple 3-second cough sounds and comeback reaction within two minutes. Generally, cough is a primary indication of over 30 medical conditions associated with non-COVID-19. By investigating morphological direction changes with dissimilarities from cough respiratory sound accuracy of 88.76% achieved in this paper, the authors collected 3-seconds cough sound data for detecting COVID-19 disease, whereas in the proposed system, cough, breath, and continuous speech are collected for analysis and detection of the COVID-19 disease and 1D CNN is implemented for classification of single input vector audio file.

In [17], Bader M et al. proposed a significant model with the combination of Mel-Frequency Cepstral Coefficients (MFCCs) and SSP (Speech Signal Processing) to extract samples from non-COVID and COVID and find the person correlation from their relationship coefficients. These findings indicate high similarity between various breathing respiratory sounds and COVID cough sounds in MFCCs, although MFCC speech is more robust between non-COVID-19 samples and COVID-19 samples. Besides these provisional findings, it is possible to remove the various patient voices with COVID-19 for future analysis. They have collected three female, four male voices from seven healthy patients, and two female, five male voices from 7 COVID-19 patients were obtained from their dataset. They were obtained COVID-19 infected patient's data from Zulekha hospital in Sharjah. The data is four times cough from each speaker, the voice of numbers counting from 1 to 10 of each speaker, and 4 to 5 times deep breath of each speaker. Besides, when recording their speech signals, the patients must sit with their heads straight in a comfortable way to extract three recordings for each speaker from smartphone devices. In this, the authors have collected very small data samples and performed all necessary operations, whereas a large dataset from Cambridge University with mutual agreement in the name of COVID-19 sounds data was collected for detecting the COVID-19 disease with desirable accuracy.

Hassan A et al. [27] implemented a system to diagnose COVID-19 positive by using the RNN model. Authors have illustrated the significant impact of RNN (Recurrent Neural Network) with the use of SSP (Speech Signal Processing) to detect the disease and specifically, this LSTM (Long Short-Term Memory) is used to evaluate the acoustic feature maps of patients' cough, breathing, and voice, in the process of early screening and diagnosing the COVID-19 virus. Compared to both coughing and breathing sound recordings, the model findings indicated poor precision in the speech test. In this work, the authors have collected a small dataset and performed the LSTM approach, and it was obtained with less accuracy. However, in our research, a large dataset was collected and is trained appropriate Deep Learning (DL) technique to achieve a better result on the COVID-19 sounds dataset.

Chaudhari G et al. [28] show that crowdsourced cough audio samples were collected worldwide on smartphones. In this process, various groups have gathered several COVID-19 cough recording datasets and used them to train machine learning models for COVID-19 detection. Each of these models was trained on data from a variety of formats and recording settings, which were collected with additional counting and vocal recordings. These datasets come from various sources, such as collecting data from clinical environments, crowdsourcing, and public media interview extraction, and are combined with COVID-19 status labels to create an AI algorithm that correctly predicts COVID-19 infection with a 77.1 percent ROC-AUC (75.2 percent–78.3 percent). In addition, without more training using the relevant samples, this AI algorithm can even generalize crowdsourced samples from Latin America and clinical samples from South Asia. In this work, the authors have collected cough data and obtained less accuracy, but the proposed approach obtains better accuracy on three respiratory sounds (human voice, breath, and cough).

Ismail MA et al. [29] proposed a model with an analysis of vocal fold oscillation to detect COVID-19; most symptomatic COVID-19 patients have mild to extreme respiratory function impairment hypothesize that through analyzing the movements of the vocal folds, COVID-19 signatures might be detectable. The authors' objective is to confirm this hypothesis and quantitatively characterize the changes observed to enable voice-based detection of COVID-19. Authors use a dynamic system model for vocal fold oscillation for this and use our recently developed ADLES algorithm to solve it to generate vocal fold oscillation patterns directly from recorded speech. Experimental findings on COVID-19 positive and negative subjects on a scientifically selected dataset show characteristic patterns of vocal fold oscillations associated with COVID-19. A data collection obtained under clinical supervision and curated by Merlin Inc., a private firm in Chile, was used for this research. The dataset contained recordings of 512 individuals who were tested for COVID-19, resulting in either positive or negative COVID-19 results. Among these, we only selected the recordings of those people who were reported within seven days after being medically examined. Only 19 citizens met this criterion. Of these, there were ten females and nine males. COVID-19 was diagnosed in 5 women and four men, and the remainder tested negative. 91.20 percent is the efficiency of logistic regression on extended vowels and their combinations. In this work, the authors have collected voice sound and predict COVID-19 with vocal sounds. In our approach, we have collected a large dataset with cough, voice, and breathing sounds.

Laguarta J et al. [30] proposed an AI (Artificial Intelligence) model from cough sound recordings to detect the COVID symptoms. This model provided a solution to pre-screen COVID-19 sound samples country-wide with no cost. It has achieved 97.1% accuracy to predict the COVID positive symptom from cough sounds, and 100% of accuracy to detect asymptomatic based on cough sounds of 5320 selected datasets is compared with chest X-ray image data [31]. Quartieri TF et al. [32] proposed a framework structure to identify COVID-19 symptomatic conditions with Signal Processing (SP) and Speech modeling techniques. This technique relied on the complexity of neuromata synchronization over speech/sound respiratory subsystem inside in the articulation, breathing, and phonation, driven by the existence of COVID symptom involving upper inflammation versus lower respiratory inflammation tract. Well-growing evidence for pre-exposure of COVID (pre-COVID) and post-COVID is provided by researcher analysis with voice meetings of 5 patients. This proposed method offered a possible capacity for flexible and continuous study to show the dynamics of patient activity in real-life settings for advanced warning and monitoring of COVID-19. In this work, the authors created one framework to identify COVID-19 symptoms but not diagnosed the exact disease. However, in our approach, COVID-19 disease is detected with good accuracy.

Sajjad A et al. proposed an end-to-end CNN approach [33] to classify environmental sounds that directly identify and describe the audio sound signal. As it breaks the signal into overlapping frames using a kernel function, the proposed solution has dealt with recorded audio sound signs of any duration. Different architectures were evaluated by considering multiple input lengths, namely initialization with a Gamma tone filter bank of the first convolutional layer for human hearing filter reflection in the labyrinth. This proposed approach has achieved around 89% accuracy by classifying urban sounds using the urban sounds 8k dataset. In this proposed approach, the authors have not used the augmentation approach and data de-noising approaches to reduce overfitting and noise reduction from the sound file, but the proposed work executed both methods to attain better audio input in order to classify appropriate respiratory sound.

Li Y et al. proposed a 1D CNN model with no attention mechanism [34] to extract the speech features from continuous speech and obtained around 88%, 65%, 75% accuracy in Emo-DB, IEMOCAP, and RAVDESS datasets. This non-attention mechanism has not given the best results in large datasets, but the proposed 1D CNN and augmentation technique with data de-noising is ready to give better and transparent results. Li F et al. proposed 1D CNN [14] for the classification of heart sound signals to extract heart sound features for attaining desired good accuracy. The 1D CNN model has been implemented for signal processing applications such as ECG classification of a patient [35,36], power electronics anomaly detection, monitoring of patient's health, and detection of fault tolerance. Salamon J et al. proposed a DNN and augmentation model [37] to classify environmental sounds and have shown a better performance with this model.

Aditya K et al. proposed CNN with a tensor deep tracking approach [38] to classify the environmental sounds and obtained 49% and 77% accuracies on the ECS-10 dataset with CNN and 56% on the ECS-10 dataset with TDSN (Tensor Deep Stacking Network). Chen X et al. proposed 1D CNN to identify flight state and obtained useful features dynamically from the basement of a newly built body wing via wind tunnel observations [39]. Pons J and Serra X proposed the CNN model for music sounds classification [40]. They have classified three sounds from the piano, drums, and flute to classify audio streaming and obtained 70% accuracy with the CNN model. Aykanat M et al. proposed the CNN model for lung sound classification [41] with collected respiratory sounds through a digital stethoscope and obtained around 80% accuracy for respiratory-based sound classification and 62% for audio-based classification.

Actually, many clinical health issues and problems (e.g., brain cancer diagnosis, prostate cancer diagnosis, etc.) are focused on Artificial Intelligence (AI) approaches. Deep learning methods can expose features of an image that are not visible in the actual image data. Primarily, CNN has been shown to be immensely helpful in the detection and training of features and has therefore been widely implemented by the scientific community. The convolutional model has been used to improves the quality of the image in low-light frames from very speed video endoscopy and was used to classify the existence of respiratory cysts through Image data, detection of pediatric TB (Tuberculosis) through respiratory X-ray images data, automatic marking of nodules during endoscopy images, cystoscopy of video image analysis. Deep learning strategies [42,43] on chest X-Rays are becoming famous with the development of deep Convolutional network and the results obtained that have been shown in various applications. In addition, a variety of data is required for the training of various learning models. The machine learning models significantly eased the method by rapidly retraining a Convolutional network with a relatively low number of images in the dataset.

In general, a number of innovative studies on the approach of classification tasks for the identification of COVID-19 from limited X-ray images and CT-scan data [44,45] were reported with promising results, but these would have to be checked on a large dataset. Some groups have upgraded or well-qualified systems to achieve better performance, while some groups use capsule networks. Mostly in this situation of a decentralized learning process, a comprehensive study on a large pulmonary disease of COVID-19 and non-COVID-19 communities is minimal and lacking. One detection model was trained to distinguish standard X-ray image data [46] and COVID, while others were trained to identify traditional images of viral infections and COVID-19 signs. In order to evaluate the effect of the necessary changes on this particular problem, both parts of the study were analyzed with and without an image improved technique. All these related studies show that there is no accurate model for diagnosing the symptoms of COVID-19 disease. So, we tend to implement the 1D CNN model and augmentation and Data De-noising Auto Encoder (DDAE) to perform better with the COVID-19 sound dataset in diagnosing COVID-19 disease, achieving better results.

## 3.    Materials and methods

### 3.1. Dataset collection

COVID-19 sounds dataset is collected from Cambridge University with mutual agreement for a research purpose. This dataset is approved at Cambridge University, Dept. of Computer Science and Tech., by adhering to the policies and regulations of the committee.

Brown C et al. [23] implemented one android app and web-based application to collect COVID-19 sounds. The main attributes of these applications are mostly similar. They have collected the past health history of a user for those who have been admitted before into the clinic. Users then entered their symptoms and reported breathing sounds (if there are any). They collected cough three times sounds, breathed heavily via their mouth 3–5 times, and read a brief statement within 30 seconds on the mobile/computer screen. Finally, they were checked for COVID-19 users and were questioned to obtain a position sample on the agreement. Besides, iOS and Android applications prompts users every two days to input additional sounds and symptoms, offering a particular chance to examine the breakthrough of sound-based patient well-being. This data is very securely encrypted in Cambridge University servers and then stored the collected data. The data was then transmitted from the telephones by connecting to Wi-Fi.

### 3.1.1.    COVID-19 sounds from the crowdsourced dataset

At the end of May 2020, University of Cambridge researchers have specifically collected around 4.5k unique from a web-based application and 2.5k samples from the android based application [23]. The authors have collected around 5k and 6k samples from different countries. Among these, around 300 users are declared COVID-19 positive patients from both web-based and android applications. The android app collects more than one sample from various users, leading to redundancies and becoming a large dataset. Hence our future work is to remove the redundancy to improve performance. They have collected and analyzed general data (past and current medical history, age, gender) along with three different sounds (vocal sound, cough, breath) from unique uses in both web-based and android applications. A dry cough is a symptom most commonly seen in this group, and the most common combination of symptoms is cough and throat. Interestingly, the most commonly affected signs are wet and dry cough, loss of ability to smell, and chest tightness (breathing) which are also considered to be the most typical combined symptoms. This is consistent with knowledge from the COVID-19 symptom monitor. The fact that coughs are one of the most documented symptoms of the COVID-19, but it also is a frequent sign of many other diseases, provides more incentive for the use of sounds as a particular symptom. Hence, the proposed 1D CNN model is to classify and diagnosis the COVID-19 from these all symptoms.

### 3.1.2. Data Augmentation

We tested five incremental sets and provided the following information. Each deformation is directly applied to the sound signal before converting it into the input data, often used to train the neural network [37]. The deceleration parameters are necessary for any increase so as to maintain the functional weight of the label. The following steps define the augmentation pair sets:

i. Stretching Time (ST): To Increase or reduce the sample sound signal (to unchanged running pitch). Based on the four factors {0.80, 0.94, 1.06, and 1.24} the duration is stretched.

ii. Shift Pitch1 (SP1): Sound/audio samples can be increased or decreased (to unchanged running pitch), and every sample can be shifted differently by four values (−1, −2, −2, −1).

iii. Shift Pitch2 (SP2): To build a second augmentation package, since our initial tests show that pitch shifting especially increases. Every sample is pitch moved by four higher values (in various sizes and shapes) this time (−2.5, −3.5, 3.5, and 2.5).

iv. Compression of Range Dynamically (CRD): These 4 parameters are compressed online with one taken from the "ICECAST" streaming server (it is an accessible software server for streaming multimedia), and three from standard Dolby E (it is a digital sound/audio stream processed by a regular stereo pair of digital sound/audio tracks).

v. Background of Noise (BN): The sample is paired with some other sequence of various kinds of audio scenes containing background noises, four sound scenes are combined for each sample (while taking respiratory sounds—environmental sound noise is combined). The mixed or combined value is generated as 'c'.

So, $c = (1 − r) \times (a + r \cdot b)$, where, a—audio signal original sample, b—background noise signal, r—random weight parameter (0.10, 0.50). Using the MUDA library, the augmentations are added, to which the reader is referred for more information on the execution of each deformation. MUDA selects the audio recording and the accompanying JAMS format annotation directory and produces the deformed audio along with the improved JAMS data containing all the deformation parameters used. In this study, actual metadata given with the audio sound dataset is imported and is utilized for evaluation into JAMS files and made accessible on the internet together with the JAMS files after deformation.

### 3.1.3. Data De-noising Auto Encoder (DDAE)

In addition, a Data De-noising Auto Encoder (DDAE) algorithm is used for the input function of the 1D convolution instead of standard input like MFCC (Mel-frequency Cepstral Coefficient) to obtain deep feature maps of respiratory sounds. The 1D CNN algorithms are then recycled to transform the input sound signal depth features to derive from the DDAE and run the pooling process. Finally, a "Softmax" classifier is used to classify the processed signal. The functionality of 1D convolution (1D CNN) is subsequently transforming the depth input features derived by the DDAE (Data De-noising Auto Encoder) method and execute pooling activity. Finally, using a Softmax classifier, the processed signals are categorized. The DDAE is derived in-depth feature respiratory sound signals by contrasted with traditional MFCC feature extraction method, the usefulness of the fundamental function of this analysis throughout the classification of respiratory sounds is thus demonstrated in section 3.2.2. The block diagram of the DDAE with 1D CNN architecture is shown in Figure 2.

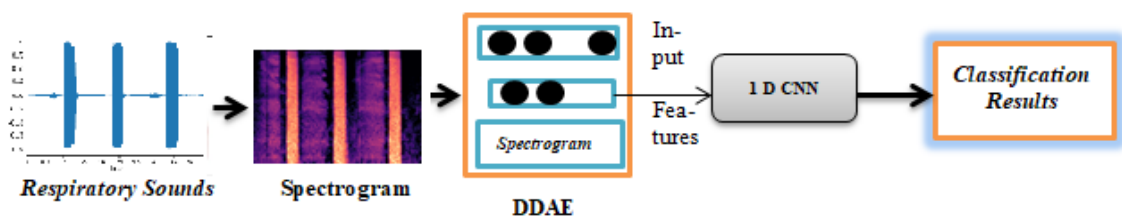

**Figure 2.** A structural architecture for proposed system with sound Data De-noising Auto Encoder.

### 3.2. Proposed 1D convolutional model

The 1D Convolutional model proposed in the present research specifically classifies respiratory sound/voice signs into usual and unusual independently of cough, voice, and breathing sounds. An augmentation-based mechanism is applied to improve the preprocessing performance of the COVID-19 sounds dataset to automatic COVID-19 disease diagnosis using the 1D convolutional network architecture. In this research, COVID-19 respiratory sound signals are extracted as input depth features by adopting of DDAE technique instead of MFCC. The proposed 1D convolutional network model is depicted in Figure 3.

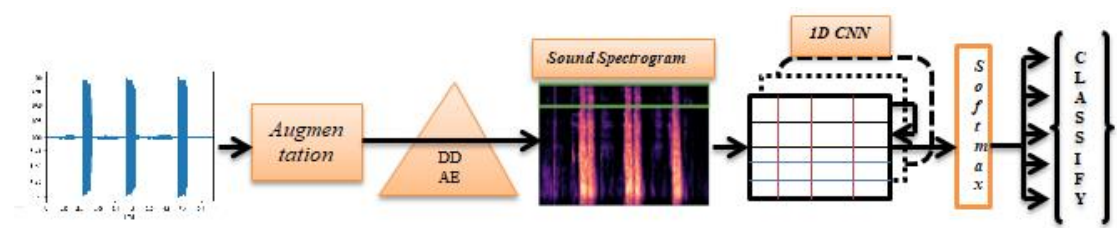

**Figure 3.** The proposed 1D Convolution Neural Network (CNN) architecture with augmentation.

### 3.2.1. Variable length of audio

There are some issues in using the 1D CNNs model for sound analysis. Indeed, the size of the input sequence must be fixed, but there exist different durations of the sound recorded from the human respiratory. Even though the COVID-19 crowdsource dataset is having a fixed size duration of each sample, but while testing, automatic sample signals may vary. Hence, the input audio length is to be fixed. Furthermore, to use it for sound signals of various lengths, it is appropriate to customize a CNN. Also, for the endless detection background noise of our respiratory input sound, CNN should be used. Splitting the voice/sound signal spectrum l into multiple buffer structures of specified size by adopting a kernel function of sufficient scope is the only way to overcome this restriction provided by the input convolutional layer. We use a dynamic size frame in our methodology to restrict the sound/voice signal to the input layer of a suggested 1D convolutional approach.

The channel size varies based on the sampling rate of the sound s signal. In addition, successive sound frames may also have a specific ratio of duplication aiming at optimizing the use of data. Ascertain pieces of the voice signals are reused; we inevitably grow the number of recordings. It is considered to be one kind of data augmentation. The frame structure and method of grouping the voice signal into suitable frames are shown in the Figure 4. In addition, the sampling frequency of respiratory sound signals input has a significant impact mainly in the depth of the input features and, ultimately, on the design model computation complexity. A great trade here between the input sound signal's performance and the operational expense of the process is assumed to be a sampling rate of 0.016 MHz for respiratory sounds.

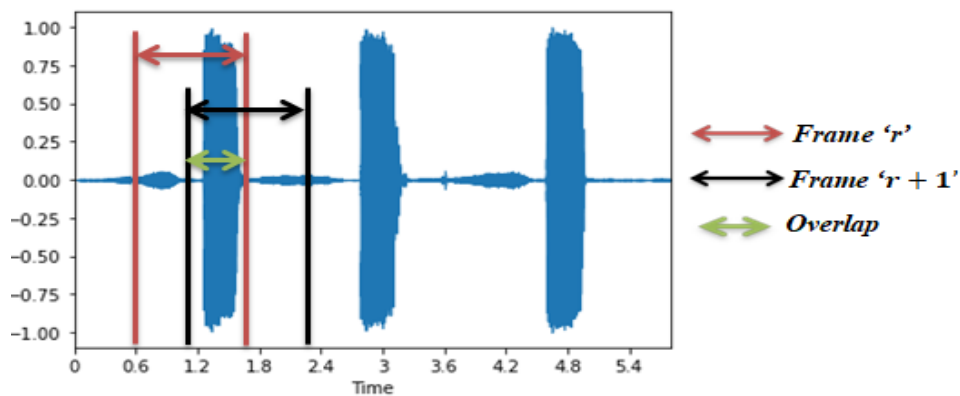

**Figure 4.** Several frames of input audio signal ($r, r + 1$) with the overlapping percentage are appropriately 50%.

### 3.2.2. 1D CNN model

The 1D CNN network flow architecture is shown in Figure 2 takes an input of the input layer kernel values as time-series signals. To extracting features and process signal layer by layer, we have used the convolutional layer and pooling layer for layer-by-layer evaluation. 1D CNN is composed of different convolutional kernels of the same size for each CNN layer. Followed by pooling layers with average pooling technique, outcomes are classified by the fully connected layer. Our model is built based on a fully connected network of 1D CNN.

The primary novel considerations of the analysis encompass:

1. Customized 1D CNN framework with input data using Data De-noising Auto Encoder (DDAE) is optimized for automated feature learning rather than individual feature extraction technique.

2. Self-adapting 1D CNN is suggested for automated parameter determination rather than depending on individual experience.

3. The DDAE will provide the input features to the 1D CNN model as a spectrum.

4. An average pooling layer accompanies the feature vector in the 1DCNN framework to readjust performance in each level in order to prevent various implications of features between training and validation results.

5. The 1D CNN architectural model can diagnosis the COVID-19 disease by analyzing human respiratory sound parameters.

The 1D Convolutional model architecture is built with a fully-connected neural network to avoid different computational sharing of the in-depth features for training data and testing data. Batch normalization is used in each 1D CNN layer to normalize each layer output. The activation function, ReLU (Rectified Linear Unit), is used for network sparsity to decrease parameter interdependence and alleviate the incidence of over-fitting problems. Six 1D convolutional layers are in the model. The function values are compressed into a single-column matrix in order to fit them into the fully-connected layer after the $6^{th}$ 1D convolutional layer output. Subsequently, the layer is flattened with two dense cells and two dropout layers, and then finally applied an activation function is "Softmax" for classification of the positive COVID-19 cough and negative COVID-19 cough, and in a similar fashion same process is applied to compute the remaining tasks. In this work, a model is constructed with six 1D CNN layers and 2-dense layers with 2-dropout cells. The new model architecture implementation process is shown in Figure 2.

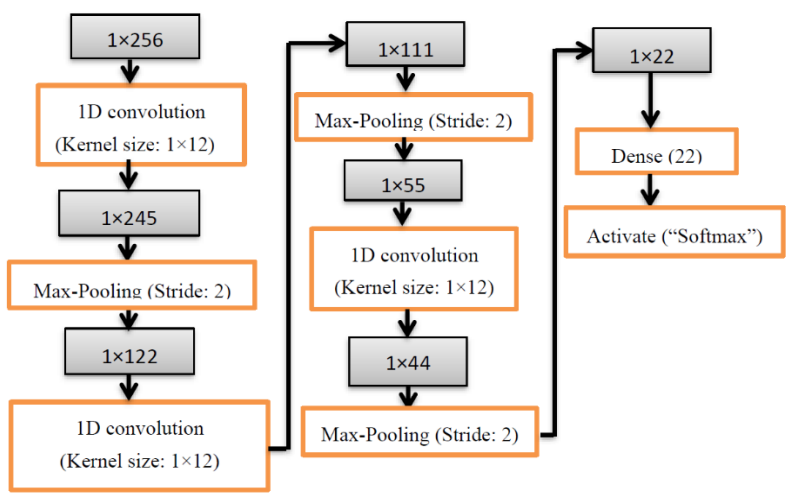

**Figure 5.** Neural network structure for the proposed 1D convolutional model.

The architectural model is constructed with a fully connected 1D convolutional network. A $1 \times 256$ dimension vector is given as an input to 1D CNN and is trained as a 2D convolutional model [47]. The model has 10-network connection features in the convolutional layer with a sigmoid transfer function. The max-pooling layer is used to generate the pooling sheet with stride 2. Finally, the Softmax classifier is used to classify each class preliminary result disease probability from COVID-19 data. The 1D CNN structure is illustrated in Figure 5 using a six-layer structure with a $1 \times 12$ kernel size. The 1D convolutional kernel sizes are extended based on column expansion in the 2D convolutional network. While preserving the convolutional filters with the same length for both networks, the outcome of various kernel sizes of a convolutional network is tested on the output of 2D and 1D Convolutional models. The length of kernel size 24 is represented in Table 1, as $1 \times 24$. The model framework is designed with the hyperparameters of six 1D CNN hidden layers, categorical cross-entropy loss function, "Adam" optimizer, 32 batch size, ReLU activation, Softmax classifier, dropout is 0.5, the number of epochs is fixed to 66, applied max-pooling after two hidden layers, and average pooling for successive hidden layers. The total trainable parameters are 87,624, and non-trainable parameters are zero. The hyperparameter objective function was updated depending on the outcomes from the first analyses, and subsequent better models were examined.

The one-dimensional convolution is similar to a standard convolutional network, except that it has raw data instead of labeled data. In order to learn a correct set of inputs, source sound data is collected across many convolution layers. According to the "local connectivity" principle, the cells in a network are connected to a particular area of the previous layer. The location of connectivity is termed the receptive field. The dataset is describing the audio sound wave by defining it as X, and it is the input to the 1D CNN output. As per functional extracting features given by $Eq.(1)$, the framework is proposed to learn a set of parameters to connect the input to the estimation Y.

$$Y = F(X|\theta) = f_k\ (.......f_2\ (f_1\ (X|\theta_1)|\ \theta_2)|\theta_k) \tag{1}$$

where the 'k' is the total number of a hidden layer of a network and the $k^{th}$ layer of the operations can be expressed as $Eq.\,(2)$,

$$Y_k = f_k(X_k|\theta_k) = h(w_j \odot X_k + \text{b}),\ ............\ \theta_k = [w_j, b] \tag{2}$$

where '⊙' is kernel operation of convolution layer and $X_k$ denoted the 'k' kernels are used for input array to extracting the features, h (·) indicates an activation operator, and b—bias function. The structure of $Y_k$, $X_k$, and $w_j$ are $(p, d)$, $(p, q)$ and $(p, d - q + 1)$ correspondingly. We have applied different pooling layers in-between each convolution layer to expand one of the following fields. The final convolution kernel layer output is flattened and used different fully connected layers are denoted as $Eq.\,(3)$:

$$Y_k = f_k(X_k|\theta_k) = h(w_j \cdot X_k + \text{b}),\ ............\ \theta_k = [w_j, b] \tag{3}$$

The output of the kernel shape is defined as mentioned in $Eq.\,(4)$, and where OL is indicated Output Length, IL defines Input Length, 'k' is the kernel filter size, 'P' indicates Padding, and 'S' defines Stride in $Eq.\,(4)$.

$$OL = \frac{IL - K + 2 \times P}{S} + 1 \tag{4}$$

The suggested 1D CNN has broad temporal information during the first convolutional network model, as it is assumed that the first layer would have a much more comprehensive view of the sound sign. Furthermore, the ambient sound signal is non-stationary about time, i.e., the amplitude or spectral quality of the signal varies. Narrower filters do not provide a general view of the spectral quality of the signal. The last pooling layer output is smoothed for all function maps and is used as an input to a fully-connected layer. After the activation feature of each convolution sheet, batch normalization is applied to decrease over-fitting [48]. Ten neurons have the last completely connected layer. Mean logarithmic error squared, as represented in $Eq.\,(5)$ is used as a function for loss ( £ ):

$$£ = \frac{1}{T} \sum_i^T \log\left(\frac{pc_i + 1}{ac_i + 1}\right)^2 \tag{5}$$

where $pc_i$, $ac_i$ are the predictive and actual classes, and 'T' is the total number of samples to calculate the loss function.

## 4. Results and discussion

We are performing three comparative experiments in this section. The first compares various kernel convolution forms, the second compares different features, and the third compares other network-layer numbers. The $F_1$ Score shows the specificity of a test in the statistical study of binary classification. The average mean value of recall & accuracy is defined as $F_1$ score, where the optimum value of the $F_1$ Score is achieved at 100% and worst at 0%. In this analysis, we use the accuracy rate & $F_1$ score for recognition to measure the method's efficiency. The following equations $[Eq.(6)$ and $Eq.(7)]$ are used to calculate the $F_1$ Score and accuracy. Where PT signifies Positive True, PF denoted Positive False, NT indicates Negative true, NF implies Negative False. Table 1 shows the accuracy and $F_1$ Score for comparison of different kernel types along with CNN kernel size and shape of the CNN. The comparison of 1D convolution and 2D convolution concerning accuracy is shown in Figure 6.

$$F_1 Score = \frac{\frac{PT}{PT+PF} \times \frac{PT}{PT+NF}}{\frac{PT}{PT+PF} \times \frac{PT}{PT+NF}} \tag{6}$$

$$Accuracy = \frac{PT+NT}{PT+NT+PF+NF} \times 100 \tag{7}$$

## 4.1. The impact of hyperparameters on results

The convolution layer in 1D CNN plays an essential role in the identification of abnormalities in the sound. The number of convolution layers for the base network model was calculated in an observational analysis by using COVID-19 crowdsourced sound dataset audio files. Sound files were divided into 13k samples, followed by 50 percent overlapping frames. 10% of the data has been used as a test dataset, 10 percent of the data was used as a validation set, and 80% data were used for training with 32 batch size for 66 epochs (the training model is saturated after 66 epochs).

**Table 1.** Different convolutional kernels comparison according to hyper-parameters (kernel size, kernel shape, kernel type) with results.

| CNN Kernel Type | CNN Kernel Size | CNN Kernel Shape | Percentage of Accuracy (%) | $F_1$Score (%) |
|---|---|---|---|---|
| 1D Convolution | 12 | $1 \times 12$ | 88.36 | 89.78 |
| 2D Convolution | | $1 \times 12$ | 86.89 | 87.24 |
| 1D Convolution | 24 | $1 \times 24$ | 89.48 | 91.26 |
| 2D Convolution | | $2 \times 12$ | 85.26 | 86.13 |
| 1D Convolution | 36 | $1 \times 36$ | 88.86 | 89.13 |
| 2D Convolution | | $3 \times 12$ | 83.13 | 84.89 |
| 1D Convolution | 48 | $1 \times 48$ | 87.96 | 88.62 |
| 2D Convolution | | $4 \times 12$ | 82.60 | 83.76 |
| 1D Convolution | 60 | $1 \times 60$ | 86.89 | 87.12 |
| 2D Convolution | | $5 \times 12$ | 80.56 | 82.67 |

The accuracy obtained by CNN 1D with one to five convolutional model test performance was 88.36%, 89.48%, 88.86%, 87.96, and 86.89%, respectively. Five 1D CNN layers are the maximum bound as the function map's minimum size on this layer was already exceeded. The same method was also used to determine the best number of convolutional layers and also their parameters for many other parameters extracted from the base model.

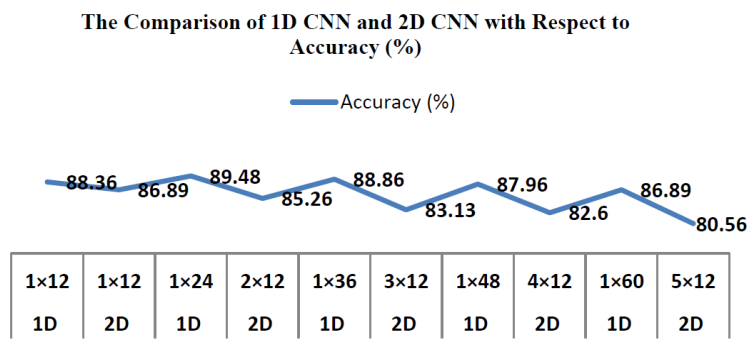

**Figure 6.** The accuracy comparison of 1D CNN and 2D CNN with different CNN kernel shapes.

Previously researchers have used the MFCC feature extraction approach to verify sound features to diagnosis COVID-19 disease. Table 2 summarizes the experimental findings. Table 2 shows that there are favorable recognition rates for the deep feature maps extracted in this analysis. Suggestions are made on the deep sound feature extracting method to extract the deep features of an audio sound file, filter the disturbance and noise in sounds, and respond favorably to the 1D convolutional network architecture features. The comparison results of the 1D convolutional model with MFCC and DDAE is based on deep sound features are depicted in Figure 7.

**Table 2.** 1D CNN features a comparison with the results.

| CNN Kernel Shape | Features Types | Percentage of Accuracy (%) | $F_1$Score (%) |
|---|---|---|---|
| $1 \times 12$ | Mel-frequency (MFCC) | 82.34 | 84.26 |
| | Deep sound features | 88.36 | 89.78 |
| $1 \times 24$ | Mel-frequency (MFCC) | 83.59 | 82.63 |
| | Deep sound features | 89.48 | 91.26 |
| $1 \times 36$ | Mel-frequency (MFCC) | 81.13 | 82.54 |
| | Deep sound features | 88.86 | 89.13 |
| $1 \times 48$ | Mel-frequency (MFCC) | 80.89 | 81.23 |
| | Deep sound features | 87.96 | 88.62 |
| $1 \times 60$ | Mel-frequency (MFCC) | 79.82 | 81.76 |
| | Deep sound features | 86.89 | 87.12 |

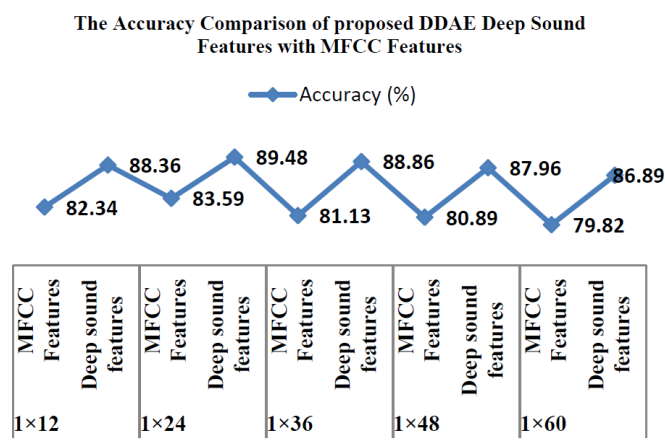

**Figure 7.** The comparison of the 1D CNN model with MFCC and deep sound features.

A proposed 1D convolutional model is proposed to achieve good performance and then joined another layer of a 1D convolutional network to design three 1D CNN layers. The pooling layers and convolution layers intermittently appear in the CNN structure. There are added pool and convolution layers. Finally, to obtain the output, the layers are completely related. The neural network output within various layers is measured. It is noticed that the network output retains almost the same when the number of hidden layers exceeds 5. The experimental results are presented in Table 3. The findings suggest that the growth in the neural network in this research enhanced the identification including accuracy of the sound signal also for various tests. The 3-layer and 5-layer 1D CNN model comparison results with different 1D CNN kernel sizes are depicted in Figure 8.

**Table 3.** The various feature type convolutional layer results in the 1D CNN model.

| CNN Kernel Shape | Features Types | Percentage of Accuracy (%) | $F_1$Score (%) |
|---|---|---|---|
| $1 \times 12$ | 3 1D CNN layers | 85.64 | 87.28 |
| | 5 1D CNN layers | 88.36 | 89.78 |
| $1 \times 24$ | 3 1D CNN layers | 86.48 | 88.89 |
| | 5 1D CNN layers | 89.48 | 91.26 |
| $1 \times 36$ | 3 1D CNN layers | 84.68 | 86.12 |
| | 5 1D CNN layers | 88.86 | 89.13 |
| $1 \times 48$ | 3 1D CNN layers | 83.96 | 85.62 |
| | 5 1D CNN layers | 87.96 | 88.62 |
| $1 \times 60$ | 3 1D CNN layers | 80.89 | 82.12 |
| | 5 1D CNN layers | 86.89 | 87.12 |

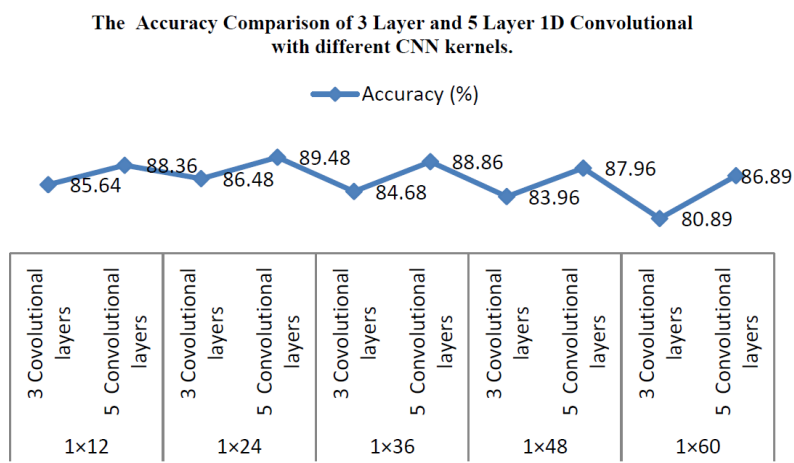

**Figure 8.** The accuracy comparison 3-layer and 5-layer 1D convolutional with different CNN kernel sizes.

The experimental findings indicate that additional information on COVID-19 sound signals is preserved by the in-depth features used for the present model by enhancing classification efficiency. The findings in Table 1 indicate that the COVID-19 sound signals used in this analysis can classify better with the proposed 1D convolutional network compared with the 2D convolutional network. This is observed when the kernel size is lower than 48 since a border kernel filter raised the performance time and progressively time space-intensive. The latter is almost 5 percent greater than that of the former in the correct rate of classification process recognition by contrasting the typical voice deep characteristics (feature maps) and MFCC characteristics (feature maps) obtained in this analysis. The deep function collected in this work is much more appropriate than the MFCC for reflecting COVID-19 sound signals. In addition, the precision of identification is immense whenever the total number of hidden convolutional layers has become enormous. Anyhow, as the number of convolutional network hidden layers increases, the sum of the estimation rises marginally, and the period of classification recognition is extended.

The collection of the deep sound features does not need preprocessing in comparison with the conventional input for any duration of COVID-19 based respiratory sounds. The respiratory sounds can thus be entered explicitly into the classification methods. Applying the 1D CNN model instead of 2D CNNs in this study is greatly enhances the ability of the entire system model to interpret COVID-19 sound signals without ever using X-ray, CT scanning, and any other reports. The proposed method achieved reliable identification efficiency based on acoustic features. The framework developed in this work is being used for regular research in everyday life or treatment modalities for the general public. If a disorder happens, then it is best to visit the clinic for more examination. This technique can only distinguish COVID-19 positive, non-COVID-19, and asthma from respiratory sounds (cough, voice sample, breath). At present, the technique could be succeeded with a limited number of samples. Therefore, in clinical practice, it has not been implemented, but it could be used for advanced detection of illness with COVID-19. This COVID-19 crowdsource data was applied in clinical practice to diagnose COVID-19 disease by Cambridge University research people [23], and with the obtained dataset from the university, an initial attempt was made to improve the performance. Further, develop the algorithms will be developed, and sample selection for clinical practice will be made in the near future.

### 4.2. Differentiating COVID-19 positive users with non-COVID-19 users from respiratory sounds

Table 4 represents the classification results of five different tasks (positive cough of COVID-19/negative cough of COVID-19, positive symptoms of COVID-19/negative symptoms of COVID-19, positive cough of COVID-19/negative COVID-19 cough with asthma, asthma breath/healthy breath, and heathy breath/cough) for ten classes (asthma breath, asthma with cough, asthma with cough and breath, breath with negative COVID-19, cough with negative symptoms of COVID-19, COVID-19 from cough, COVID-19 from breath, healthy symptoms from breath, heathy symptoms from cough, COVID-19 positive cough + breath). The classification network was developed that classified five tasks from 10 class categories. The first task represents binary classification used for discriminating whether the user is a COVID-19 positive or non-COVID-19 user. The second task discriminates positive cough of COVID-19 from negative cough symptoms of COVID-19. Task 3 is for discriminating positive cough symptoms of COVID-19 from negative COVID-19 asthma cough. Task 4 for discriminating users having asthma breath or healthy breath and the last task is meant for discriminating normal cough from asthma cough. The exact accuracy comparison for classification results of five different tasks for respiratory sounds is shown in Figure 9. Table 5 displays the comparison results for different tasks for the proposed 1D CNN model with the existing SVM classifier. In total, around 2%, more accuracy is achieved than existing methods. The accuracy comparison of the proposed model with different tasks and the previous SVM model [1] is shown in Figure 10.

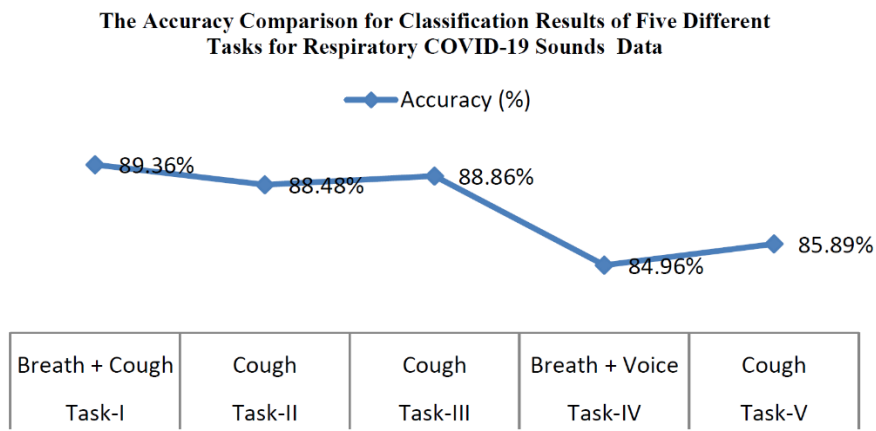

**Figure 9.** The accuracy comparison for classification results of five different tasks for respiratory COVID-19 Sounds Data (Task-1: Positive COVID-19/Negative COVID-19; Task-2: Positive cough of COVID-19/Negative cough of COVID-19; Task-3: Positive cough of COVID-19/Negative COVID-19 Asthma Cough; Task-4: Asthma Breath/Healthy Breath; Task-5: Asthma cough/Normal Cough).

**Table 4.** Classification Results of five tasks for respiratory COVID-19 sounds with previous models.

| Task | Modality | Accuracy (%) | F1Score (%) |
|------|----------|--------------|-------------|
| Positive COVID-19/Negative COVID-19 | Breath + Cough | 89.36 | 90.78 |
| Positive cough of COVID-19/Negative cough of COVID-19 | Cough | 88.48 | 89.26 |
| Positive cough of COVID-19/Negative COVID-19 Asthma Cough | Cough | 88.86 | 89.13 |
| Asthma Breath/Healthy Breath | Breath + Voice (Single sentence voice) | 84.96 | 86.62 |
| Asthma cough/Normal Cough | Cough | 85.89 | 87.12 |

*4.3. The comparison of proposed model with exiting model on COVID-19 crowdsourced sound data*

The data benchmarks show that there are some discriminatory indications in the data while testing for COVID-19 user coughs mixed with breathing may be a good predictor. Specifically, test accuracy for task 1 is around 90% by combining breath sounds with cough sounds. Task 2 and task 3 achieve around 88% of accuracy. We can diagnose asthma with breath, voice, and cough with an accuracy of around 85%. The 1D CNN is a proposed model for classification for COVID-19 disease from the user samples. Data set is obtained from Cambridge University on mutual agreement. The data has a fixed number of fewer samples (13k). In the near future, more samples will be collected to improve the performance with a multi-level deep convolutional model. A plan will be made to implement a multi-layer deep convolutional model to improve the performance of the diagnosis of COVID-19 disease. Table 6 depicts the comparative analysis of the proposed model with previous audio classification models on different respiratory sound data in recent years.

**Table 5**. Comparison of proposed model with previous models.

| Model | Dataset | Accuracy |
|---|---|---|
| SVM + PCA [23] | COVID-19 Crowdsourced Data | Task-I: 80% |
| | | Task-II: 82% |
| | | Task-III: 80% |
| SVM with Augmentation [23] | COVID-19 Crowdsourced Data | Task-II: 87% |
| | | Task-III: 88% |
| 1D CNN with Augmentation and DDAE (Proposed) | COVID-19 Crowdsourced Data | Task-I: 90% |
| | | Task-II: 88% |
| | | Task-III: 88% |
| | | Task-IV: 84% |
| | | Task-V: 86% |

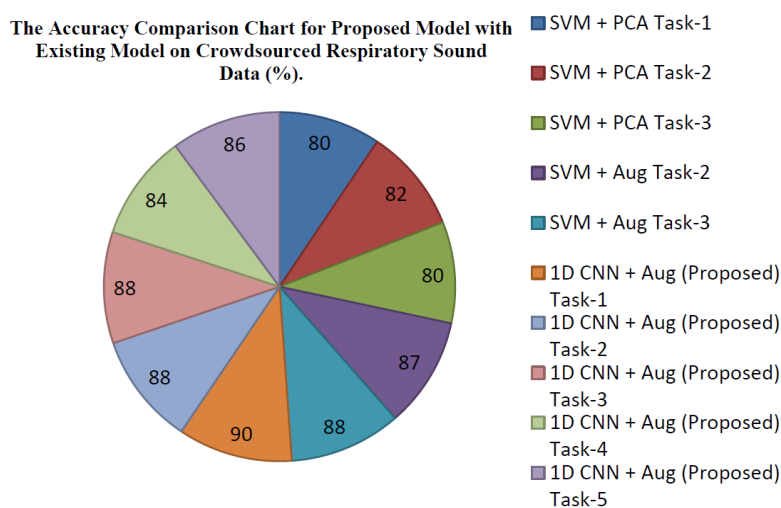

**Figure 10.** The accuracy comparison chart for the proposed model (1D CNN with augmentation & DDAE) with five different tasks (Task-1: Positive COVID-19/Negative COVID-19; Task-2: Positive cough of COVID-19/Negative cough of COVID-19; Task-3: Positive cough of COVID-19/Negative COVID-19 Asthma Cough; Task-4: Asthma Breath/Healthy Breath; Task-5: Asthma cough/Normal Cough) and previous models with three different tasks (Task-1: Positive COVID-19/Negative COVID-19; Task-2: Positive cough of COVID-19/Negative cough of COVID-19; Task-3: Positive cough of COVID-19/Negative COVID-19 Asthma Cough).

The detection of sound features is not required for preprocessing compared to the traditional method for any duration of the respiratory sound. Therefore, the respiratory sound can be specifically entered into the classification networks. In this analysis, the implementation of the 1D Convolutional model network rather than the 2D convolutional model greatly enhances the opportunity of the entire system model to interpret COVID-19 respiratory sound signals. The proposed model will obtain accurate identification output based on respiratory sound signal features without the use of reference X-ray or CT scan images. The framework developed in this work may be used for normal everyday life assessments or medical centers for the general public. If a disorder occurs, it is essential to visit the clinic for any further testing. Restricted by the sample size obtained, this approach can only identify respiratory disorders such as asthma, COVID-19 positive, and COVID-19 negative symptoms at current. Thus, it has never been used in clinical practice but can be used for early detection of respiratory disease. In the future, we will further develop algorithms and sample selection for use in clinical practice.

**Table 6.** The comparative analysis of the proposed model with previous audio classification models on different respiratory sound data in recent years.

| Model | Dataset | Modality | Accuracy (%) |
|---|---|---|---|
| DCT + MFCC [28] | Clinical Dataset 1 | Cough | 58.60 |
| | Clinical Dataset 1 | | 71.80 |
| MFCC + ResNet50 [30] | MIT open voice Data | Cough + Voice | 79.2 |
| DCT + MFCC [17] | COVID-19 Sample sound data collected from Android App. | Cough + Breath + Voice | 79 |
| BI_at_GRU [3] | Sample data Collected from Ruijin Hospital | Breath + Thermal Video | 83.69 |
| DTL-MC [4] | COVID-19 Sample sound data collected from Android App. | Cough + Voice | 88.76 |
| SVM + PCA [23] | COVID-19 Crowdsourced Data | Cough + Breath + Voice | Task-I: 80% |
| | | | Task-II: 82% |
| | | | Task-III: 80% |
| VGG Net [23] | COVID-19 Crowdsourced Data | Cough + Breath + Voice | Task-II: 87% |
| | | | Task-III: 88% |
| 1D CNN with Augmentation and DDAE (Proposed) | COVID-19 Crowdsourced Data | Cough + Breath + Voice | Task-I: 90% |
| | | | Task-II: 88% |
| | | | Task-III: 88% |
| | | | Task-IV: 84% |
| | | | Task-V: 86% |

## 5. Conclusions

Artificial Intelligence (AI) based models entered into the real-world to diagnosis the COVID-19 symptoms from human-generated sounds such as voice/speech, cough, and breath. The CNN (Convolutional Neural Network) model was used to solve many real-world problems with Artificial Intelligence (AI) based machines. In this work, we proposed a 1D (one-dimensional) CNN (Convolutional Neural Network) to diagnose COVID-19 disease with human respiratory sounds collected from the COVID-19 sounds crowdsourced dataset. A Data De-noising Auto Encoder (DDAE) was adopted to extract the acoustic sound signals in-depth features instead of traditional MFCC. As a result, around 4% accuracy is achieved than traditional MFCC. We have classified COVID-19 sounds, asthma sounds, and regular healthy sounds using a 1D CNN classifier and shown around 90% accuracy to detect the COVID-19 disease from respiratory sounds. The proposed model improves efficiently to classify COVID-19 sounds for detecting COVID-19 positive symptoms. We will use the deep convolutional model with multi-feature channels on this COVID-19 crowdsource sounds dataset to enhance better performance in future work.

## Acknowledgments

We would like to express our sincere gratitude to Prof. Cecilia Mascolo, clinical scientists at Cambridge University, for sharing the dataset. We acknowledge everyone who is trying to stop the COVID-19 pandemic.

## Author contributions

All authors have participated in (a) conception and design, or analysis and interpretation of the data; (b) drafting the article or revising it critically for important intellectual content; and (c) approval of the final version. The authors have not been submitted this article too nor under review to any journal or publishing body.

## Conflict of interest

The authors have declared no conflict of interest.

## References

1. World Health Organization. Coronavirus disease 2019 (covid-19). (2020) Available from: https://www.who.int/.
2. Wang Y, Hu M, Li Q, et al. (2020) Abnormal respiratory patterns classifier may contribute to large-scale screening of people infected with COVID-19 in an accurate and unobtrusive manner. *arXiv:2002.05534 [cs.LG]*.
3. Jiang Z, Hu M, Lei F, et al. (2020) Combining Visible Light and Infrared Imaging for Efficient Detection of Respiratory Infections Such as Covid-19 on Portable Device. *arXiv:2004.06912 [cs.CV]*.
4. Imran A, Posokhova I, Qureshi HN, et al. (2020) AI4COVID-19: AI enabled preliminary diagnosis for COVID-19 from cough samples via an app. *Inform Med Unlocked* 20: 100378.
5. Shuja J, Alanazi E, Alasmary W, et al. (2020) COVID-19 open source data sets: a comprehensive survey. *Appl Intell* 21: 1–30.
6. Rasheed J, Jamil A, Hameed AA, et al. (2020) A survey on artificial intelligence approaches in supporting frontline workers and decision makers for the COVID-19 pandemic. *Chaos Solitons Fractals* 141: 110337.
7. Alafif T, Tehame AM, Bajaba S, et al. (2021) Machine and Deep Learning towards COVID-19 Diagnosis and Treatment: Survey, Challenges, and Future Directions. *Int J Environ Res Public Health* 18: 1117.
8. Ritwik KVS, Shareef BK, Deepu V. (2020) Covid-19 Patient Detection from Telephone Quality Speech Data. *arXiv:2011.04299v1 [cs.SD]*.
9. Kranthi KL, Alphonse PJA (2021) A literature review on COVID-19 disease diagnosis from respiratory sound data. *AIMS Bioeng* 8: 140–153.
10. Huang Y, Meng S, Zhang Y, et al. (2020) The respiratory sound features of COVID-19 patients fill gaps between clinical data and screening methods. *medRxiv 2020.04.07.20051060*.

11. Shi J, Zheng X, Li Y, et al. (2018) Multimodal Neuroimaging Feature Learning With Multimodal Stacked Deep Polynomial Networks for Diagnosis of Alzheimer's Disease. *IEEE J Biomed Health Inform* 22: 173–183.

12. Brabenec L, Mekyska J, Galaz Z, et al. (2017) Speech disorders in Parkinson's disease: early diagnostics and effects of medication and brain stimulation. *J Neural Transm (Vienna)* 124: 303–334.

13. Erdogdu SB, Serbes G, Sakar CO (2017) Analyzing the effectiveness of vocal features in early telediagnosis of Parkinson's disease. *PLoS ONE* 12: e0182428.

14. Li F, Liu M, Zhao Y, et al. (2019) Feature extraction and classification of heart sound using 1D convolutional neural networks. *EURASIP J Adv Signal Process* 59.

15. Klára V, Viktor I, Krisztina M (2011) Voice Disorder Detection on the Basis of Continuous Speech, In: Jobbágy Á, *5th European Conference of the International Federation for Medical and Biological Engineering*. IFMBE Proceedings, Springer, Berlin, Heidelberg.

16. Verde L, De Pietro D, Sannino G (2018) Voice Disorder Identification by Using Machine Learning Techniques. *IEEE Access* 6: 16246–16255

17. Bader M, Shahin I, Hassan A (2020) Studying the Similarity of COVID-19 Sounds based on Correlation Analysis of MFCC. *arXiv:2010.08770 [cs.SD]*.

18. Sahidullah Md, Saha G (2012) Design, analysis and experimental evaluation of block-based transformation in MFCC computation for speaker recognition. *Speech Commun* 54: 543–565.

19. Srinivasamurthy RS (2018) Understanding 1D Convolutional Neural Networks Using Multiclass Time-Varying Signals. *All Thesis*. Available from: https://tigerprints.clemson.edu/cgi/viewcontent.cgi?article=3918&context=all_theses.

20. Zhao W, Singh R (2020) Speech-Based Parameter Estimation of an Asymmetric Vocal Fold Oscillation Model and its Application in Discriminating Vocal Fold Pathologies. *ICASSP 2020– 2020 IEEE International Conference on Acoustics, Speech and Signal Processing (ICASSP)*, Barcelona, Spain.

21. Kumar A, Gupta PK, Srivastava A (2020) A review of modern technologies for tackling COVID-19 pandemic. *Diabetes Metab Syndr* 14: 569–573.

22. Deshpande G, Schuller B (2020) An Overview on Audio, Signal, Speech, & Language Processing for COVID-19. *arXiv:2005.08579 [cs.CY]*.

23. Brown C, Chauhan J, Grammenos A, et al. (2020) Exploring Automatic Diagnosis of COVID-19 from Crowdsourced Respiratory Sound Data. *Proceedings of the 26th ACM SIGKDD International Conference on Knowledge Discovery & Data Mining*.

24. Han J, Qian K, Song M, et al. (2020) An Early Study on Intelligent Analysis of Speech under COVID-19: Severity, Sleep Quality, Fatigue, and Anxiety. *arXiv:2005.00096v2 [eess.AS]*.

25. Orlandic L, Teijeiro T, Atienza D (2020) The COUGHVID crowdsourcing dataset: A corpus for the study of large scale cough analysis algorithms. *arXiv:2009.11644v1 [cs.SD]*.

26. Singh R (2019) Production and Perception of Voice. In: *Profiling Humans from their Voice*. Springer, Singapore.

27. Hassan A, Shahin I, Alsabek MB (2020) COVID-19 Detection System using Recurrent Neural Networks. *2020 International Conference on Communications, Computing, Cybersecurity, and Informatics (CCCI)*, Sharjah, United Arab Emirates.

28. Chaudhari G, Jiang X, Fakhry A, et al. (2021) Virufy: Global Applicability of Crowdsourced and Clinical Datasets for AI Detection of COVID-19 from Cough. *arXiv. PPR: PPR272849*.

29. Ismail MA, Deshmukh S, Rita S (2020) Detection of COVID-19 through the Analysis of Vocal Fold Oscillations, *arXiv:2010.10707v1 [eess.AS]*.

30. Laguarta J, Hueto F, Subirana B (2020) COVID-19 Artificial Intelligence Diagnosis Using Only Cough Recordings. *IEEE Open J Eng Med Biol* 1: 275–281.

31. Wang L, Lin ZQ, Wong A. (2020) COVID-Net: a tailored deep convolutional neural network design for detection of COVID-19 cases from chest X-ray images. *Sci Rep* 10: 19549.

32. Quartieri TF, Talker T, Palmer JS (2020) A Framework for Biomarkers of COVID-19 Based on Coordination of Speech-Production Subsystems. *IEEE Open J Eng Med Biol* 1: 203–206.

33. Sajjad A, Patrick C, Alessandro LK (2019) End-to-end environmental sound classification using a 1D convolutional neural network. *Expert Sys Appl* 136: 252–263.

34. Li Y, Baidoo C, Cai T, et al. (2019) Speech Emotion Recognition Using 1D CNN with No Attention. *2019 23rd International Computer Science and Engineering Conference (ICSEC)*, Phuket, Thailand.

35. Serkan K, Onur A, Osama A, et al. (2019) 1D Convolutional Neural Networks and Applications: A Survey. *Mech Sys Signal Proc* 151: 107398.

36. Kiranyaz S, Ince T, Abdeljaber O, et al. (2019) 1-D Convolutional Neural Networks for Signal Processing Applications. *ICASSP 2019–2019 IEEE International Conference on Acoustics, Speech and Signal Processing (ICASSP)*, Brighton, UK.

37. Salamon J, Bello JP (2017) Deep Convolutional Neural Networks and Data Augmentation for Environmental Sound Classification. *IEEE Signal Proc Lett* 24: 279–283.

38. Aditya K, Deepak G, Nguyen NG, et al. (2019) Sound Classification Using Convolutional Neural Network and Tensor Deep Stacking Network. *IEEE Access* 7: 7717–7727.

39. Chen X, Kopsaftopoulos F, Wu Q, et al. (2019) A Self-Adaptive 1D Convolutional Neural Network for Flight-State Identification. *Sensors* 19: 275.

40. Pons J, Serra X (2019) Randomly Weighted CNNs for (Music) Audio Classification. *ICASSP 2019–2019 IEEE International Conference on Acoustics, Speech and Signal Processing (ICASSP)*, Brighton, UK.

41. Aykanat M, Kılıç O, Kurt B, et al. (2017) Classification of lung sounds using convolutional neural networks. *J Image Video Proc*, 65.

42. Ismael AM, Abdulkadir S (2021) Deep learning approaches for COVID-19 detection based on chest X-ray images. *Expert Sys Appl* 164: 114054.

43. Minaee S, Abdolrashidi A, Su H, et al. (2021) Biometrics Recognition Using Deep Learning: A Survey. *arXiv:1912.00271 [cs.CV]*.

44. Yazdani S, Minaee S, Kafieh R, et al. (2020) COVID CT-Net: Predicting Covid-19 from Chest CT Images Using Attentional Convolutional Network. *arXiv:2009.05096 [eess.IV]*.

45. Jain R, Gupta M, Taneja S, et al. (2021) Deep learning-based detection and analysis of COVID-19 on chest X-ray images. *Appl Intell* 51: 1690–1700.

46. Khan AI, Shah JL, Bhat MM (2020) CoroNet: A deep neural network for detection and diagnosis of COVID-19 from chest X-ray images. *Comput Methods Programs Biomed* 196: 105581.

47. Wu Y, Yang F, Liu Y, et al. (2018) A Comparison of 1-D and 2-D Deep Convolutional Neural Networks in ECG Classification. *arXiv:1810.07088v1 [cs.CV]*.

48. Ioffe S, Szegedy C (2015) Batch Normalization: Accelerating the deep network training by reducing internal covariate shift, Proceedings of the 32$^{nd}$ International Conference on Machine Learning. *Proceed Mach Learn Res* 37: 448–456.

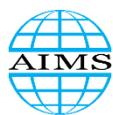 AIMS Press